\documentclass[pdflatex,sn-mathphys-num]{sn-jnl}

\usepackage{graphicx}%
\usepackage{multirow}%
\usepackage{amsmath,amssymb,amsfonts}%
\usepackage{amsthm}%
\usepackage{mathrsfs}%
\usepackage[title]{appendix}%
\usepackage{xcolor}%
\usepackage{textcomp}%
\usepackage{manyfoot}%
\usepackage{booktabs}%
\usepackage{algorithm}%
\usepackage{algorithmicx}%
\usepackage{algpseudocode}%
\usepackage{listings}%

\theoremstyle{thmstyleone}%
\theoremstyle{thmstyletwo}%

\theoremstyle{thmstylethree}%

\begin{document}

\title[Evaluating LLM-based AI agents integrated with materials synthesis tools]{Evaluating LLM-based AI agents integrated with materials synthesis tools: the case of atomic layer deposition}


\author[1*]{\fnm{Angel} \sur{Yanguas-Gil}}\email{ayg@anl.gov}

\affil[1]{\orgdiv{Applied Materials Division}, \orgname{Argonne National Laboratory}, \orgaddress{\street{9700 S Cass Ave}, \city{Lemont}, \postcode{60062}, \state{Illinois}, \country{USA}}}


\abstract{
This review provides an overview of the different strategies that can be used to evaluate the performance of AI models and agents based on large language models (LLMs) 
for materials synthesis. After providing a brief overview of the key technologies behind the current generation of AI agents based
on LLMs, we summarize the different approaches to evaluating these models in the context of materials science and in particular on materials synthesis, with a specific emphasis
on scenarios in which the models are directly integrated with experimental tools. We discuss evaluation strategies spanning knowledge and reasoning benchmarks,
tool-use benchmarks, and closed loop benchmarks involving the interaction with experimental systems or realistic virtual tools.
We use atomic layer deposition (ALD) as a case study, emphasizing how existing
approaches in the literature both build from general approaches used beyond materials science and can be generalized to other materials synthesis techniques. Finally, we provide a practical evaluation framework to evaluate LLMs in the context of materials synthesis}




\maketitle

\section{Introduction}\label{sec:intro}

\begin{figure}[ht]
\centering
\includegraphics[width=0.5\textwidth]{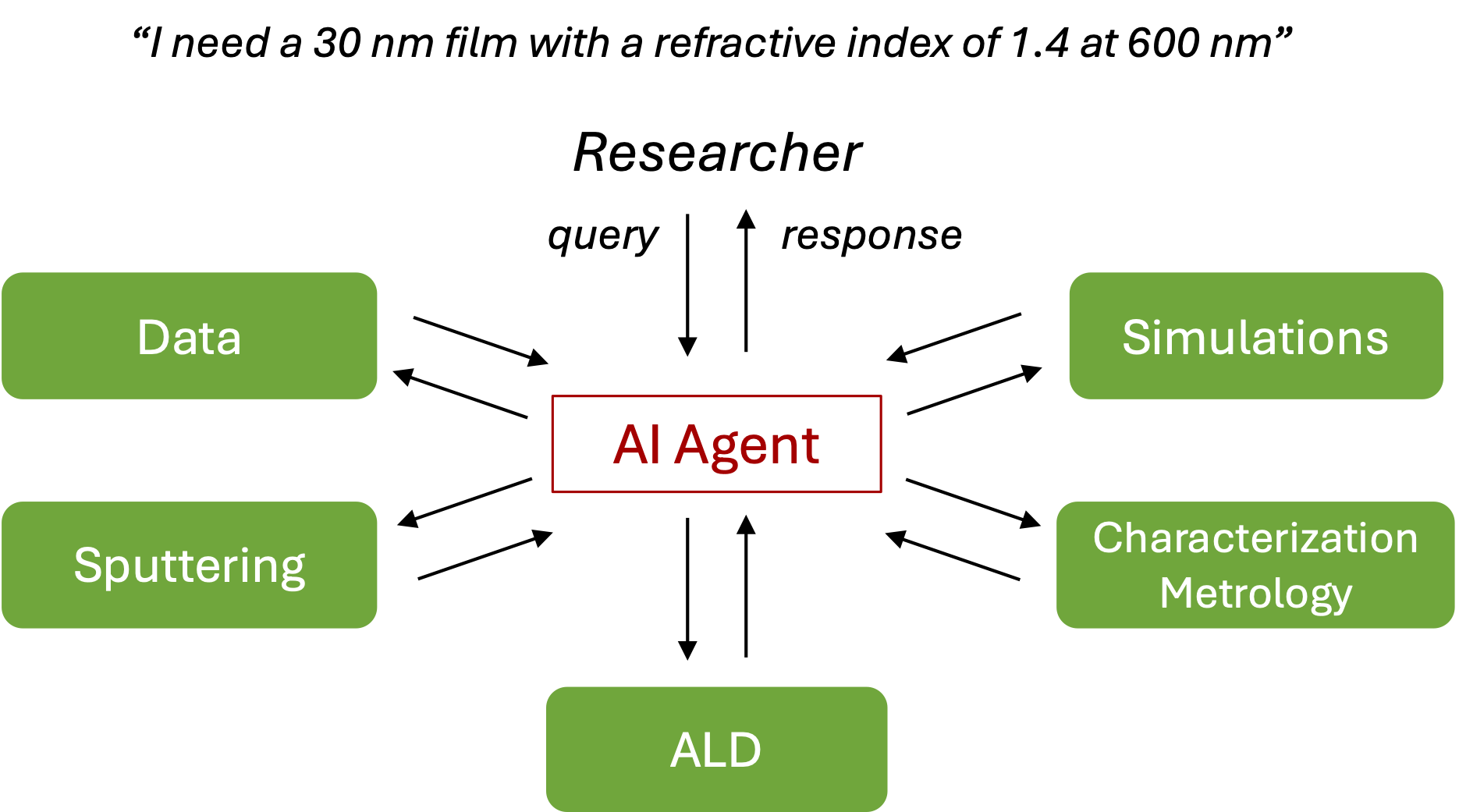}
\caption{AI agents powered by LLMs can potentially orchestrate multiple complex research workflows involving the interaction with multiple materials synthesis, characterization,
and simulation tools as well as databases and prior data}\label{fig:agent}
\end{figure}

In recent years, there has been growing interest in integrating generative AI models and agents with materials synthesis tools to realize autonomous platforms for
materials discovery and process optimization\cite{Hase2019,MacLeod_selfdriving,Stach_review_2021,Paulson2021}. The underlying vision is that agents powered by the most recent generation of large language
models (LLMs) could orchestrate complex research workflows\cite{Boiko2023} involving multiple synthesis and characterization tools, simulations and prior data (Fig. \ref{fig:agent}), much as they 
already excel in tasks such as code generation.
Unlike conventional machine learning (ML) models designed for specific tasks, LLM-based agents can potentially address a much broader set of tasks, interpreting natural-language
instructions, accessing and searching for data, and managing multi-step workflows involving both simulations and experiments.

Current efforts to integrate AI with experiments fall into two broad categories: the development of end-to-end complex autonomous platforms mentioned above and the augmentation of existing instruments with AI agents. Examples of the latter include characterization techniques such as atomic force microscopy and thin-film growth techniques such as pulsed laser deposition and atomic layer deposition\cite{MandalAFM2025,Harris_PLD,Yanguas-Gil2026aa}. While the ability to augment or retrofit existing equipment and tools allows us to tap into the wealth of experimental tools in materials science labs and consequently have a broad impact, it has its own unique challenges, including how to design effective interfaces to experimental equipment that were not originally designed for automation, and how to make these interfaces compatible with various types of AI/ML algorithms, which can range from deterministic space searches to the latest generation of LLM-enabled agents.

In this review, we focus on the problem of how to quantify the performance of AI agents integrated with experimental tools. Evaluating LLM-based agents is critical to understand
how well they will perform in real research tasks, identify existing limitations, and quantify the actual value of AI integration. The importance of benchmarks and evaluation is bound to increase as LLMs and AI agents progress to higher technology readiness levels and are consequently reused for multiple projects and applications. This is something that we are observing today, with commercial and open-weight LLMs being actively explored for different research tasks\cite{Zhang:2025aa}.
In the context of materials synthesis, the cost and time associated with running experiments and the risk of inadvertently damaging valuable equipment make quantifying the performance of AI models particularly relevant. The stakes are even higher in manufacturing settings such as semiconductor processing, where the impact of cross-contamination or simply low yields can propagate downstream over multiple processing steps before being detected.

Developing evaluation strategies and benchmarks to help us understand the capabilities and shortcomings of AI agents and models is therefore crucial to pre-empt any problems and derisk their integration with experiments. This is a responsibility that is shared between AI researchers and domain experts: AI researchers make extensive use of benchmarks to quantify model performance during the training or fine-tuning of large language models\cite{Chang:2024aa}, but most of the benchmarks are meant to be run in environments that are isolated from experimental labs. Materials scientists
can play a critical role in the development of targeted benchmarks to ensure that the models can be evaluated in scenarios that are relevant for downstream uses in materials research.

This review has therefore two goals: to provide an overview of how experimental tools are integrated with LLM-based agents, and to summarize strategies used for evaluating the
 performance of these agents. This work approaches the subject from a materials science perspective. The emphasis is not so much on the training of new models as on how we can evaluate the performance of these models and agents. In Section \ref{sec:llm_intro} we provide a quick overview of LLM-based technologies, how they have been leveraged for materials science and related scientific applications, and the main evaluation strategies used in the literature. In Section \ref{sec:llm_eval}, we review the evaluation strategies used to understand how well LLMs perform for materials science, and for
 materials synthesis in particular. We focus on atomic layer deposition as a model system to explore how some of the main strategies used for LLM evaluation are particularized for materials synthesis. Finally, in Section \ref{sec:discussion} we conclude with some recommendations, including a practical evaluation framework for materials synthesis.
 
\section{LLM-based agents and their integration with experiments}
\label{sec:llm_intro}

\subsection{Overview of LLMs, multimodal LLMs, and AI agents}
\label{sec:overview}

Large language models are a class of deep learning models trained on large corpora of text to predict the probability distribution over the next token in a sequence. A key
breakthrough was the introduction of the transformer architecture, which provided a self-attention mechanism capable of representing long-range correlations present
in complex texts\cite{vaswani2023attentionneed}. The generative pre-trained transformer (GPT) family is arguably the best-known example\cite{Brown_GPT_2020}.

A defining characteristic of LLMs is the probabilistic nature of their outputs: given a prompt, the model samples from a learned probability distribution over possible continuations, producing outputs that are stochastic even for identical inputs. This stochasticity can have important implications for applications in science and in materials synthesis in particular, where
reproducibility and consistency are critically important as these technologies mature from research artifacts to real-world use, particularly when integrated with experimental and simulation capabilities.

Applications built on top of LLMs typically rely on an orchestration layer, or harness: this is a software layer that manages the interaction between the model and the user. For instance, the harness is what gives the perception of having a conversation with a model. After each new user input, the harness sends the model the conversation history up to that point, therefore priming the LLM to provide a response built on top of the conversation history.

Two key capabilities developed using a combination of model training and harnesses are multi-step reasoning strategies and tool use. In the context of multi-step reasoning,
strategies such as chain-of-thought\cite{Wei_chain_2022} and self-consistency sampling\cite{wang2023selfconsistency} have powered a new generation of reasoning LLMs that are capable of solving complex logical problems.
Examples of such reasoning models include OpenAI's o3 as well as open-weight models such as DeepSeek-R1\cite{Guo_2025} and Qwen\cite{Yang2025}. These techniques, which also require
their own specific training strategies, rely on breaking down queries into intermediate reasoning steps that are addressed by LLMs before a final answer is returned to the user.

In tool use, or function calling, LLMs are trained to generate structured output indicating which tool to use when their prompt includes a structured description of available resources \cite{Schick_toolformer_2023,patil2023gorillalargelanguagemodel} (for instance a Python interpreter or a database query interface). The harness then interprets the model output, runs any tool as needed, and generates a new prompt that is passed to the model with the response. This model-tool-updated prompt cycle can run many times before the information is sent back to the user. 

One limitation of conventional LLMs is that they are limited to text. Multimodal LLMs (MLLMs) are designed to overcome this limitation, and they are capable of ingesting one or more additional modalities (image, video, audio, etc.) besides text. MLLMs contain other pre-trained components to encode the main features of the additional input modalities. They also have
a cross-modal connector that bridges the output of the different encoders with the language model. For instance, the LLaVA architecture uses a multilayer network to project features encoded by a pre-trained vision encoder into the language embedding space\cite{Liu2023}. The model is trained using a visual instruction-tuning methodology\cite{Liu2023}.

AI agents based on LLMs leverage all these individual tools to both reason about complex, multi-step tasks and interact with tools and their environment\cite{yao2023_reasoning}. The concept of an \emph{agent} predates the rise of LLMs \cite{Russel_AI_book}, and it broadly refers to systems that interact with their environment to achieve concrete goals. A key difference of LLM-based agents is that they do not require task-specific training or programming.

The usefulness of tool use has led to the development of higher-level standard protocols and practices that can be used to integrate experimental tools with directly with harnesses or agents. One of the most common examples is the so-called Model Context Protocol (MCP), where tools are interfaced through servers that can be accessed by agents or harnesses\cite{Du:2026aa}. The agent has access to a list of tools, each with their own API and documentation, which are used as needed. Another example is the use of skills, plain text configuration files that provide instructions of how to leverage specific resources. As in the MCP case, a list of skills are made available to the agent and used as needed. The main advantage of these higher-level approaches for the materials scientist is that they allow sidestepping harness engineering altogether. The domain scientist can focus instead on the design of an interface that can be integrated with off-the-shelf harnesses or agents.

\subsection{Integrating LLM-based agents with experimental tools}
\label{sec:llm_experiments}

At a high level, integrating LLM-based agents with experimental tools typically involves the generation of API calls or other structured instructions. In this section, we will exemplify this through four representative examples in the literature:

A first representative example is Coscientist, one of the earliest frameworks to demonstrate an LLM-powered agent capable of autonomously designing and executing chemical experiments\cite{Boiko2023}. In its original design, Coscientist comprises an LLM-planner that has access to four different commands: {\sc google},
{\sc python}, {\sc documentation}, and {\sc experiment}. A simple harness uses the LLM to generate calls to any of the four modules as needed or until a maximum number of iterations is reached. Interpretation and execution are handled by an automation layer that does not rely on LLMs. The results of these experiments are then sent back to the harness and used to generate subsequent calls.

A second example is the integration of an atomic force microscopy (AFM) tool\cite{MandalAFM2025}. Mandal et al. implement two agents: an AFM Handler Agent that is used for experimental control and a Data Handler Agent for analysis. The interface with the experimental system
is done through a Python-based API, enabling direct control of the experimental system. The AFM Handler Agent has control of a code-execution tool, which directly runs
Python code generated by the LLM. This allows the generation of complex sequences of instructions through Python scripts. 

\begin{figure}[ht]
\centering
\includegraphics[width=0.5\textwidth]{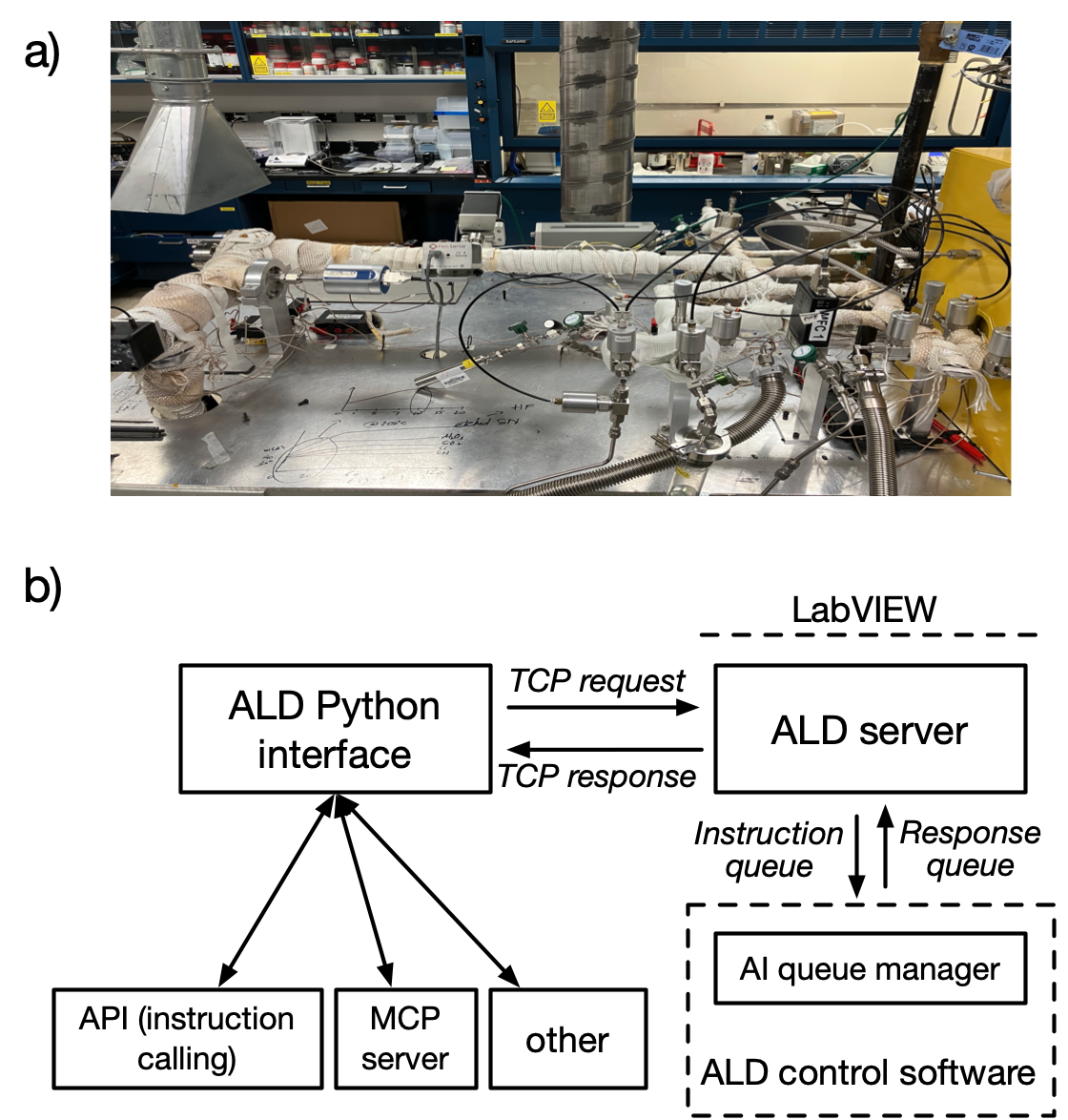}
\caption{Example of LLM integration with an atomic layer deposition tool: (a) Picture of the cross-flow ALD reactor, including in-situ ellipsometer; (b) Scheme of the interface design, where the LabVIEW software controlling the ALD tool is accessed by a Python API via TCP sockets. This API allows multiple types of integration with both conventional
ML algorithms and LLMs, including through MCP servers or direct tool use. Figure adapted from Ref. \cite{Yanguas-Gil2026aa} with permission.\label{fig:alddesign}}
\end{figure}

The third example is the integration of an LLM-based agent with an atomic layer deposition tool (Fig. \ref{fig:alddesign})\cite{Yanguas-Gil2026aa}. In this case, a translation layer
sits between the outward-facing Python API, which defines a small set of operations, and the LabVIEW backend, which is used to operate the tool. This Python API is designed to allow for three types of integration with LLMs: through tool use, the Model Context Protocol (MCP), and a custom harness similar to the execution loop of the Coscientist architecture
from Boiko et al.\cite{Boiko2023}. In contrast to the AFM case, in the specific example reported in Ref. \cite{Yanguas-Gil2026aa}, the LLM outputs text in structured format, in this case JSON, that is then parsed and translated into a specific tool call.

A final fourth example showcases the integration of experimental tools via MCP servers in the context of a synchrotron imaging beam line\cite{Du:2026aa}: Du \emph{et al}'s 
agent architecture integrates tool-augmented actions with multimodal reasoning, and optional long-term memory to support both autonomous
procedures and interactive user-guided measurements. Their Experiment Automation Agents are also vision-capable, which allows the authors to carry out tasks such as 
automated zone plate focusing or search for features in images using natural language.

\begin{figure}[ht]
\centering
\includegraphics[width=0.5\textwidth]{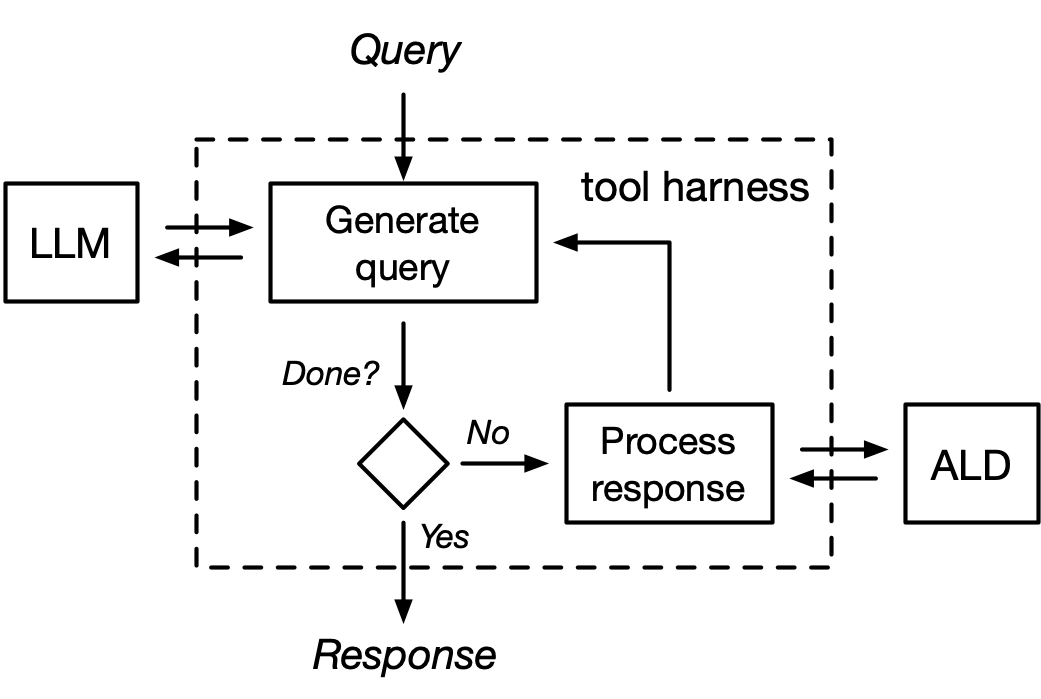}
\caption{A high-level structure of a simple LLM agent interacting with an experimental tool: the LLM-based agent comprises a harness that implements a query-response-action loop until
the LLM considers that the task is done or a maximum number of iterations has taken place. This harness is responsible for translating the model response into the instructions that are used to communicate with various experimental tools and resources available to the agent. Examples such as Coscientist\cite{Boiko2023} or agents used for ALD process optimization share a similar top-level architecture.}
\label{fig:harness}
\end{figure}

While there are other examples in the literature, these four cases showcase one of the most common design patterns to integrate LLMs with experimental tools: 
\begin{enumerate}
\item LLMs generate tool calls in some sort of structured format, such as JSON, Python or an application-specific schema.
\item A harness translates model outputs into executable actions, often with an iterative control loop. The exact structure of the agent can change from case to case and involve multiple modules and even multiple agents each specialized in a part of the project. This harness may include non-LLM modules such as parsers, safety checks, schedulers, or conventional ML components.
\item A query-response-action loop is common, especially when the agent must incorporate feedback from tools, databases, or experiments before deciding on the next step.
\end{enumerate}
Figure \ref{fig:harness} shows the high-level design of a simple LLM agent interacting with experimental tools that incorporates these key ideas.

\section{Evaluating LLMs and LLM-based agents in materials science}
\label{sec:llm_eval}

In this section, we summarize the main evaluation strategies relevant to materials science and materials synthesis. Section \ref{sec:eval_general}  provides an overview of evaluation strategies for LLMs in scientific applications and materials science.
Section \ref{sec:ald} then focuses on evaluation for materials synthesis through the case study of atomic layer  deposition.

\subsection{General evaluation strategies and benchmarks in materials science}
\label{sec:eval_general}

\subsubsection{Evaluating knowledge, reasoning, and text generation}
\label{sec:eval_base}

As shown in the prior section, regardless of whether the output is produced directly by the LLM or through a harness with some sort of internal logic, LLMs are designed to generate
output text from an input prompt. At the highest level, both general-purpose and materials-science-specific LLM benchmarks comprise a collection of questions or inputs together
with a scoring method for individual responses and the overall score for the benchmark. These benchmarks can be run either directly calling the model or through an evaluation harness that standardizes the evaluation process. Most of the relevant examples are designed to be run by directly calling the model via an API without the need to use a harness. 

To evaluate knowledge and reasoning capabilities, LLM benchmarks typically use a few standard approaches: multiple choice questions (MCQ) are a common approach, being extensively used as part of general purpose and materials science and scientific benchmarks. A useful aspect of MCQs is that evaluation can be carried out either by directly computing the
probability for each answer option after inputting the question, or by evaluating an LLM-as-judge approach, in which an open-ended response is evaluated by an LLM that determines whether it corresponds to the right answer. This ease of evaluation and of encoding the right response makes MCQs  one of the most common approaches for early benchmarks. One such example is GPQA\cite{GPQA}, whose main set comprises 448
expert-level questions in biology, physics, and chemistry. These questions were designed by domain experts to require a significant amount of domain-specific reasoning and to
resist retrieval through internet searches. Within the context of materials science, MCQs are part of benchmarks such as MaScQA\cite{Zaki_MaScQA_2024}, which includes more than 350 MCQs comprising 14 subdomains within materials science, from materials manufacturing to electrical properties and mechanical behavior.
Finally, the structured nature of MCQs is ideally suited to explore automatic question generation\cite{Li:2025aa}.

Questions requiring open-ended responses are used when probing the ability to generate numerical values, for instance to evaluate the model's ability to retrieve physical properties
of materials or to provide the solutions to numerical problems. They are also used to answer questions requiring the generation of text in specific format, such as chemical formulas or SMILES or even to generate code. They are evaluated using a combination of rule-based and LLM-based evaluations. For instance, MatSciBench is a benchmark comprising 1340 questions curated from materials science textbooks comprising 1236 numerical questions and 104 formula questions\cite{MatSciBench}. The authors evaluate the answers
using primarily a rule-based approach based on parsing the output using regular expressions based on that used to evaluate LLM's answers to math questions. Another example
is ChemBench, which integrates MCQs with open response questions in the field of chemistry\cite{chembench}.
Benchmarks requiring free-text generative outputs typically use either LLM-as-a-judge scoring approaches or human Likert-scale ratings. One example of the second case is
ALDBench, which was developed to evaluate the ability of LLMs to reason about atomic layer deposition\cite{YanguasGil2025}. This will be described in detail in Section \ref{sec:aldbench}.

Finally, recent benchmarks have started to incorporate modalities beyond texts to evaluate the potential of multimodal models. MaCBench is a benchmark that focuses on evaluating the model performance on research tasks\cite{MaCBench}. Each entry in MaCBench comprises pairs of images and text-based questions. They employ a rule-based parser that uses regex-based functions to handle both MCQs and numerical values. As in the case of MatSciBench, an LLM extractor is used to handle cases where the parser fails. Circi et al 
evaluated the ability of LLMs to understand tables in materials science, creating a dataset of 2,512 entries and comparing the performance of using image vs text (OCR and structured format) inputs on data extraction tasks\cite{Circi_MatTables_2024}.

\begin{figure}[ht]
\centering
\includegraphics[width=7cm]{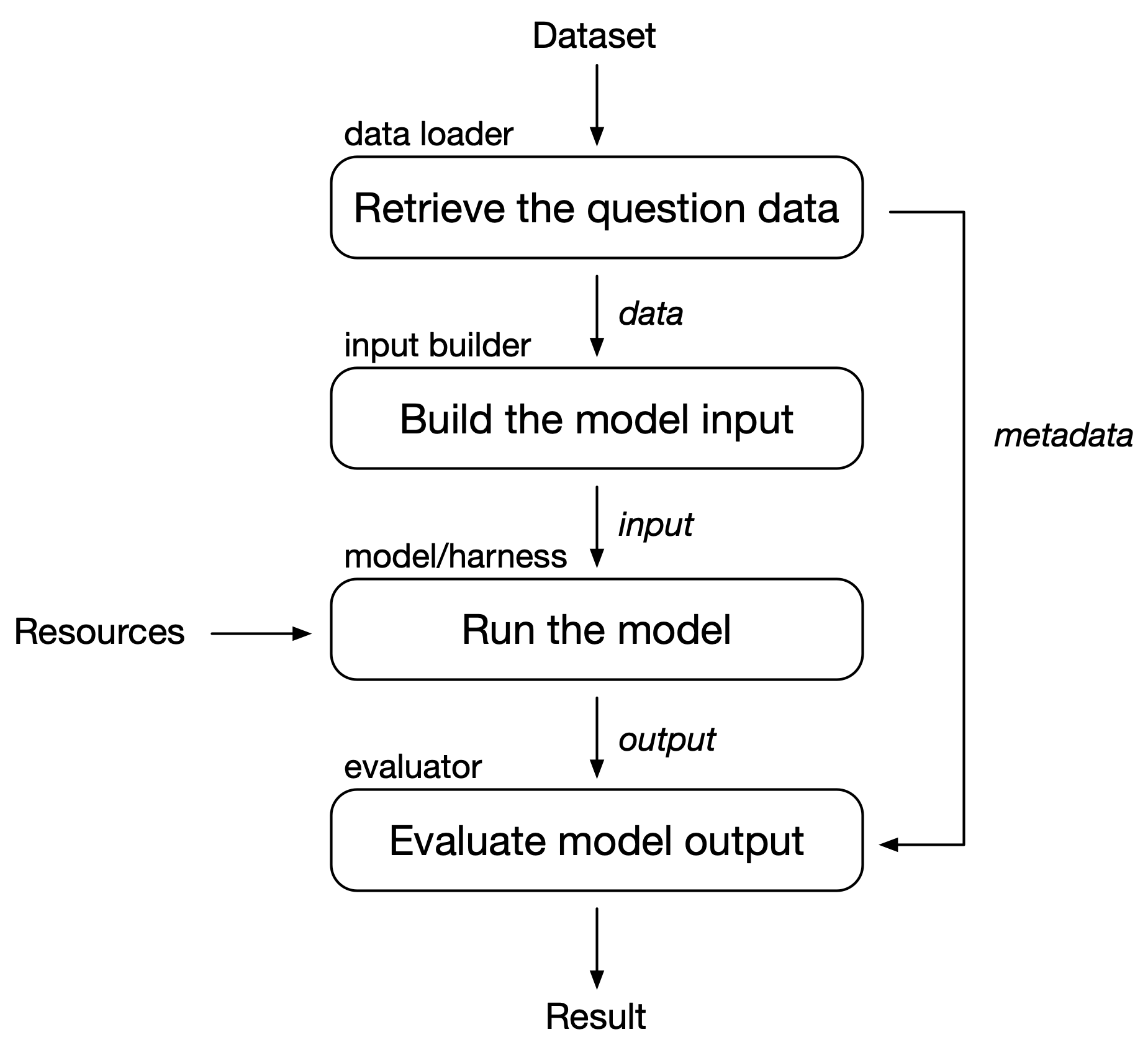}
\caption{Basic LLM evaluation pipeline: for each question in the benchmark, a \emph{data loader} takes data and the \emph{input builder} transforms it into a properly formatted input
that is then passed to the model. The \emph{data loader} can also extract any relevant metadata, such as the difficulty or the topic of the question. The output of the model is then passed through an 
\emph{evaluator} to extract the results and compute the model score.\label{fig:eval_workflow}}
\end{figure}

Despite their diversity, these benchmarks share a common structure: 1) a \emph{dataset} containing the information required to build each input to the model; 2) a \emph{builder} function that transforms each entry into a model input, often a prompt and, for multimodal benchmarks, additional modalities such as images or spectra; 3) an \emph{evaluator} function that  parses the model output and compute a score. The resulting evaluation pipeline is summarized in 
Figure \ref{fig:eval_workflow}. The data for each question may include relevant metadata, such as the difficulty, subcategory or additional context for the evaluator. In some cases, additional resources are made available to the model, such as the ability to access the open internet or a collection of tools. These components can be implemented in a few lines of Python code, shared as their own code repositories, or be part of complex testing harnesses that leverage online repositories to download models and the data required to run a specific benchmark\cite{eval-harness}.

\subsubsection{Evaluating tool use and interactive agent behavior}
\label{sec:eval_tool}

\begin{figure}[ht]
\centering
\includegraphics[width=7cm]{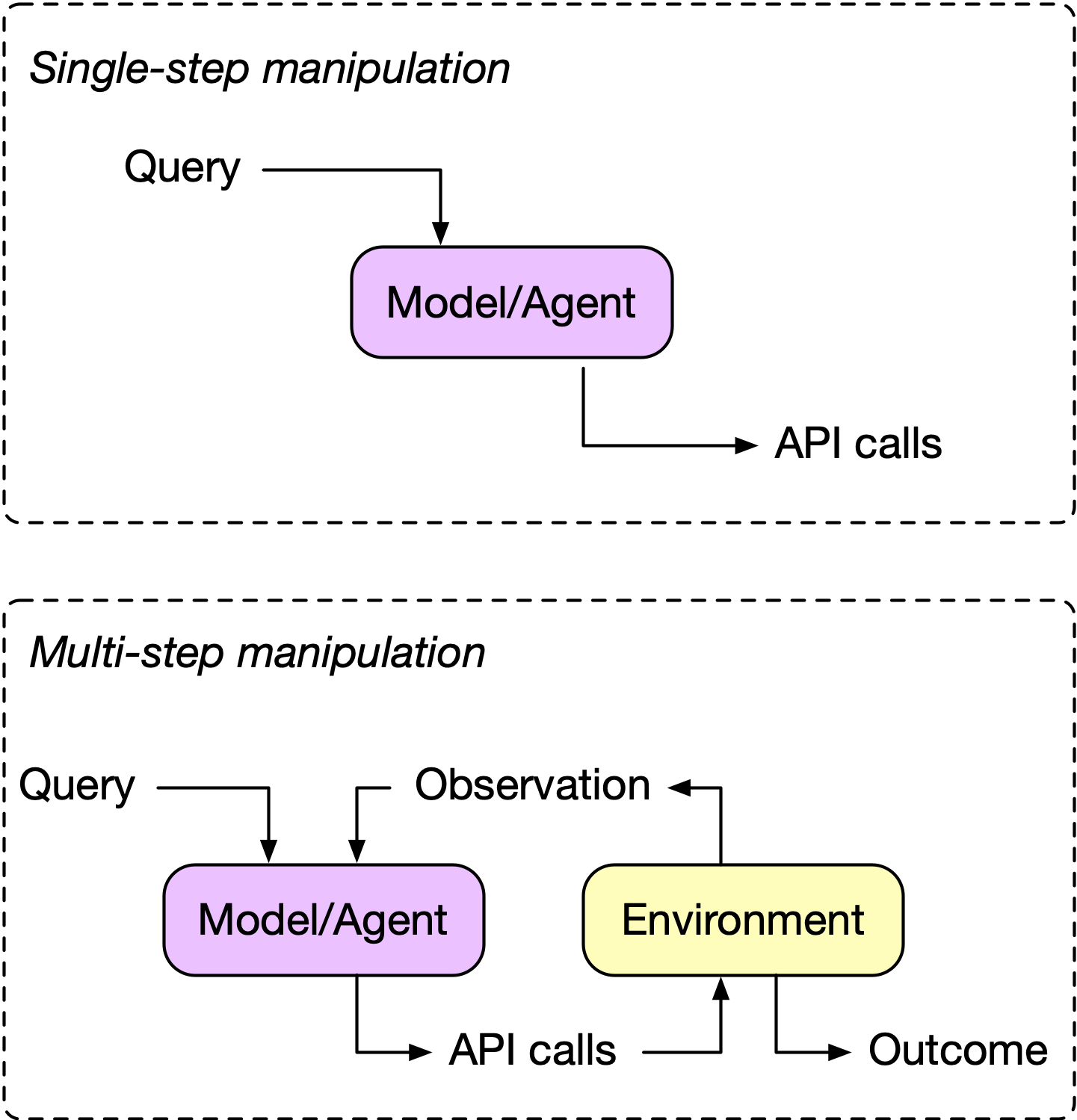}
\caption{Tool manipulation examples fall into two broad scenarios: in the single step manipulation case, the model or agent responds with a sequence of API calls; in the
multi-step manipulation scenario, the agent interacts with the environment, receiving feedback to a sequence of tool calls until a desired outcome takes place.\label{fig:eval_tool}}
\end{figure}

As tool use has become an increasingly important application of LLMs, a number of benchmarks have emerged that evaluate the ability of AI agents and models to interact
with external tools. Some of these general purpose benchmarks provide conceptual models that are useful in the context of materials science.
For instance, in the work from Xu et al introducing the ToolBench benchmark\cite{toolbench}, the authors provide a simple conceptual model for tool use comprising two different scenarios (Figure \ref{fig:eval_tool}) in the \emph{single step manipulation scenario}, a model or agent has to generate one or more calls to external tools to accomplish a specific goal. Models can be evaluated using the general evaluation pipeline described in Section \ref{sec:eval_base} (Figure \ref{fig:eval_workflow}). In contrast, in the \emph{multistep manipulation scenario},  the model or agent must interact iteratively with an environment, generating tool calls and incorporating the environment's feedback until it reaches an exit step.

This framework can be directly applied to the integration of AI with scientific tools. From a materials science perspective, this has a series of advantages. First, the ToolBench framework
treats the model or AI agent as a black box, being broadly defined as a system that takes a text prompt as input and executes a series of actions. Second, communication takes
place through an API or interface. This interface is the natural place where the experiment interfaces with the AI model. Consequently, the design of benchmarks that are specific to
a given tool requires thinking about the interface.

The single step manipulation scenario computes the accuracy by comparing the function calls with the desired function call. This means that it is possible to evaluate model without having to physically connect them to the instruments. The benchmark can therefore be the first step in understanding the feasibility of using AI to drive a specific tool. This provides
a practical way to iterate on interface design and improve the performance of the model or agent before connecting it to a real piece of equipment.

The evaluation of the multistep generation scenario requires the interaction with a tool or an environment\cite{yao2023_reasoning,liu2025agentbenchevaluatingllmsagents}. However, as we show in Section \ref{sec:eval_aldopt}, in some cases it is possible to replace the real tool with a virtual tool that is an approximation of a real system. Even if that system is not tied to a specific tool, since the main use case of generative AI is its applicability to a wide range of conditions, a model's ability to solve challenges even in an idealized case can provide insights on the expected performance when connected to a real system.

\subsubsection{Benchmarks for materials synthesis}

While some general-purpose and scientific benchmarks do contain materials synthesis queries, materials synthesis is not covered in depth in chemistry and materials science benchmarks. For instance, MaScQA includes 91 questions on materials manufacturing and 35 questions on materials processing, focused primarily on metallurgy\cite{Zaki_MaScQA_2024}.  MatSciBench focuses on materials processing also in the context of metallurgy\cite{MatSciBench}. Overall, most materials science-specific benchmarks do not address materials synthesis in an explicit way (see, for instance, Ref. \cite{Miret:2025aa} for a perspective on LLMs for materials science).

Likewise, the number of benchmarks that address specific materials synthesis techniques is also still very limited.
In addition to the ALD case addressed in Section \ref{sec:ald} \cite{YanguasGil2025,Yanguas-Gil2026aa,Yanguasgil_reasoning},  
there are only a few examples in the literature. For instance, Cho et al developed a twenty open-response question benchmark to evaluate the performance of retrieval augmented generation (RAG) systems using the specific case of graphene synthesis. These questions were created by domain experts and evaluated using four different methods, a metric of similarity with the ground truth based on BERTscore, LLM and human judges, and the automated RAG evaluation framework RAGAS\cite{Cho_RAG_2026}. 

In the broader context of inorganic chemistry, Kim et al explored LLM's ability to predict the synthesizability of chemical compounds\cite{Kim_LLMinorg_2024}. Out of 393,053 
unique compositions contained in the Open Quantum Materials Database and the Materials Project, they defined as synthesizable the subset with references in the Inorganic Crystal Structure Database (40,817). The structure was similar to the MCQ approach, with the models being asked to generate if the compound was synthesizable or not based on the chemical formula. They also developed a benchmark that asked the model to identify precursors for materials synthesis based on 11,923 unique reactions and 311 precursors based on
the synthesis dataset from Kononova et al\cite{Kononova_minedreactions_2019}. In this case, each question was open ended and the response was parsed against a predefined template and the correct list of precursors.

Overall, the current state of the literature points to significant gaps that need to be addressed if we want to fully understand the capabilities and limitations of LLMs and LLM-based
agents in materials synthesis.

\subsection{Atomic layer deposition as a case study}
\label{sec:ald}

The benchmarks described in Section \ref{sec:eval_general} have helped illuminate the potential and limitations of existing LLMs in the context of material science and the approaches to evaluate the performance of LLMs and agents. While not trying to be exhaustive, many of the benchmarks mentioned in Section \ref{sec:eval_general} provide a great starting point to understand the potential of LLMs. They also play a key role in AI research, assisting in the development of a new generation of models specialized for materials science research.

However, there is still an important gap between the general capabilities evaluated by general purpose benchmarks such as MatSciBench or ChemBench, and the level of granularity required to evaluate models meant to be deployed along with materials synthesis tools to assist in the research process. This has led to the development of more specialized benchmarks that target specific subdomains and tasks.

In this section we use atomic layer deposition (ALD) as a model system to understand how to apply different evaluation strategies to materials synthesis. The examples in
this section adapt the general evaluation ideas introduced in Sections \ref{sec:eval_base} and \ref{sec:eval_tool} and particularize them for this thin film growth technique. They provide a great case study that can inform how to approach evaluation of models and agents in the context of materials synthesis.

\subsubsection{Atomic layer deposition}
\label{sec:aldintro}

Atomic layer deposition is a thin film growth technique  based on the self-limited reactions of a precursor and a co-reactant with surfaces. In contrast to chemical vapor deposition,
which is predominantly a continuous process, in ALD exposures to precursor and co-reactant are separated by purge times to prevent their simultaneous presence in the reactor. An ALD cycle involving one precursor and one co-reactant is defined by four time intervals: $t_1$, $t_2$, $t_3$, $t_4$, where $t_1$ and $t_3$ are the precursor and co-reactant dose times, 
respectively, and $t_2$ and $t_4$ are the precursor and co-reactant purge times. A repetition of multiple cycles produces an ALD growth. 

The optimization of an ALD process typically involves identifying the shortest dose and purge times that still achieve saturated growth per cycle. If the dose times are too small, the growth per cycle is lower than the saturated growth per cycle. If the dose times are too long, precursor is wasted and the process takes longer. An ALD reactor frequently has multiple precursor and co-reactant channels. Executing a growth therefore involves picking a set of precursors and co-reactant, setting conditions such as the reactor temperature, and the right dose and purge times. ALD can be extended to more complex processes in which different precursors are alternated to grow ternary, alloyed, or doped materials. It may also use surface functionalization steps to achieve growth selectivity or modulate the saturated growth per cycle. There are a number of reviews in the literature that explore the range of ALD processes and applications in depth\cite{Kessels2025}.

ALD is a particularly useful model system for evaluating LLMs in materials synthesis. First, its high degree of automation makes it a natural candidate for integration with both conventional
ML methods and LLM-based agents. Second, it allows evaluation at multiple levels of granularity,  from identifying relevant chemistries and selecting available precursors in a tool, to carrying out closed-loop process optimization.

In the following sections, we will introduce three different methodologies for evaluating LLM-based models and agents in the context of atomic layer deposition: an open response benchmark meant to be evaluated by human experts, a benchmark for the use of ALD experimental tool, and a benchmark for ALD process optimization based on reasoning LLMs.

\subsubsection{ALDBench: an expert-reviewed open-response benchmark}
\label{sec:aldbench}

\begin{figure}[ht]
\centering
\includegraphics[width=7cm]{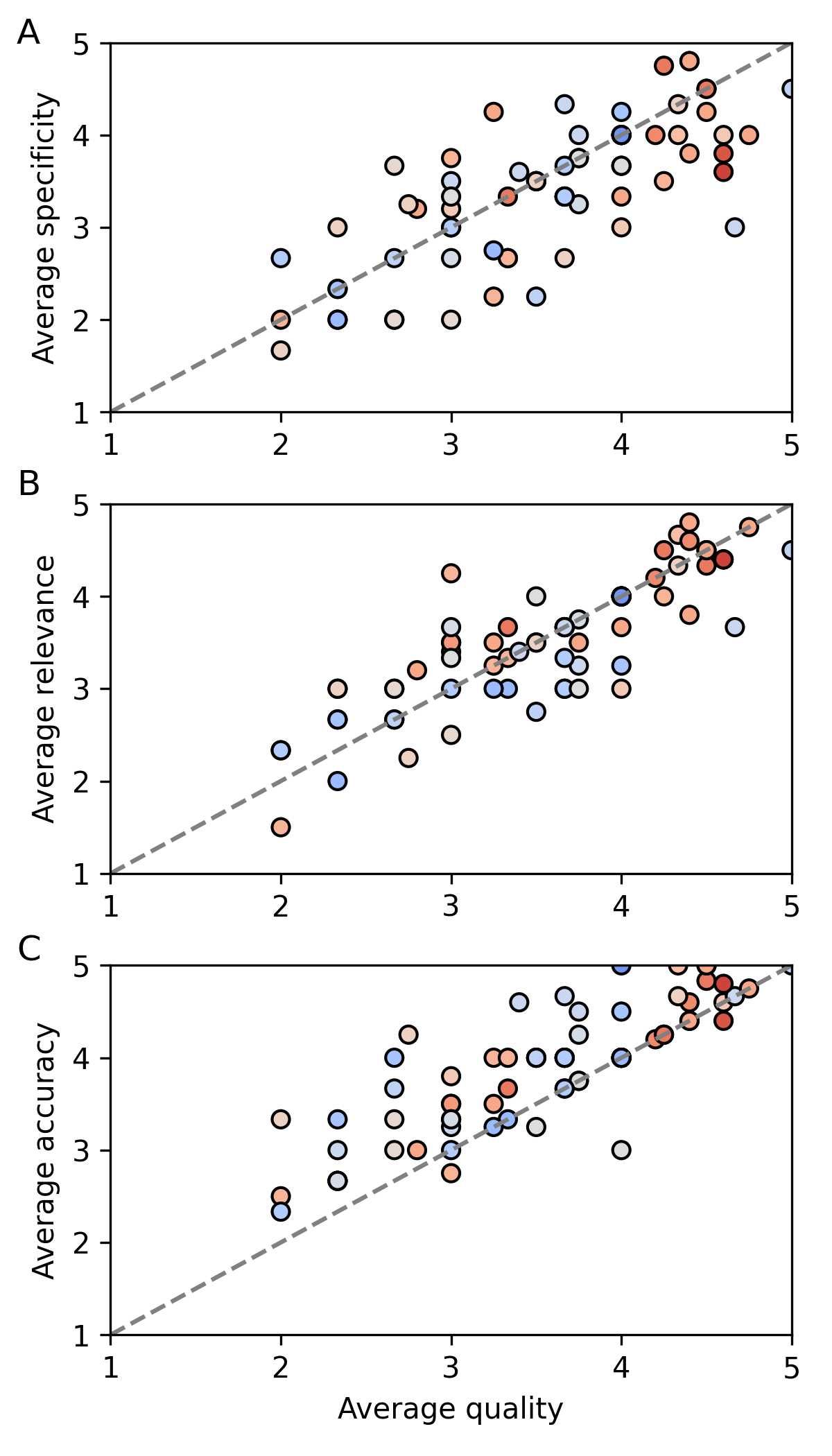}
\caption{Correlation between the average quality of the GPT4o responses to the ALDBench benchmark and (A) the specificity of the response; (B) the relevance of the response; (C) the accuracy of the response. Points are colored by average difficulty of each question (red-easy, blue-hard). Reproduced from Ref. \cite{YanguasGil2025} with permission. \label{fig:eval_open}}
\end{figure}

ALDBench provides an example of an open response benchmark developed specifically to understand LLM's capabilities in the context of materials synthesis\cite{YanguasGil2025}. ALDBench
comprised 70 questions grouped into four categories: "how to grow", "specific questions about ALD processes", "general ALD knowledge", and "applications". 
LLM responses were graded by seven human experts across four criteria: Quality, Specificity, Relevance, and Accuracy, using a 1 to 5 Likert scale. Questions were
also graded in terms of their Difficulty (1-Easy, early graduate, 5-Hard, top expert) and Specificity (1-General, 5-Specific, quantitative). This design made it possible both to compute scores for each category and explore correlations among question and response attributes. An example is shown in Figure \ref{fig:eval_open} showing the correlation between the average quality and the other three criteria. Relevance, in particular, tracks more closely with the perception of quality of the human reviewers.

Grading both questions and answers also allowed exploring correlations between each of the dimensions. The analysis split the scores into two groups, one for high scores and another one for below average, and built contingency tables for all possible pairs of different criteria. Through the application of the Fisher exact test to each contingency table allowed them to identify significant differences (p-values $<$ 0.05). The results were consistent with the presence of statistically significant correlations for three different cases: the overall quality of the response and the difficulty of the question (higher difficulty led to lower quality responses), the relevance of the response and the difficulty of the question (more difficult questions lead to less relevant response), and the accuracy of the response and the specificity of the question (more specific questions led to less accurate answers).

A key advantage of using open ended responses is that experts can apply their expertise to study in detail the model responses, for instance to find evidence of hallucination, implausible recommendations or overly vague but reasonably sounding responses, all of which are described in Ref. \cite{YanguasGil2025}. In fact, while three models, GPT-3.5, GPT-4.0 and GPT-4o were tested, only GPT-4o produced responses of high enough quality to merit a dedicated study. On the other hand, the use of human experts as reviewers limits the scale at which these experiments have been run. This has led to the exploration of model-as-a-judge approaches.

\subsubsection{Evaluating tool use in atomic layer deposition}
\label{sec:eval_aldtool}

As mentioned in Section \ref{sec:llm_experiments}, Ref \cite{Yanguas-Gil2026aa} presents the design of an experimental atomic layer deposition chamber that has been integrated with an AI agent. At the core of that
integration is a simple API that provides a small number of functions that models can use to request information and execute growths. In order to understand the performance
of the AI agents, the work presents a systematic comparison of a common agent architecture using different underlying LLMs.

The benchmark comprised a series of challenges that each required a single tool use call, in this case the call required to execute a given growth. For each challenge, a query was passed to the model together with the required background information, which in this case comprised the ALD tool configuration and potentially a list of prior experiments. Similar to the approach used
in ToolBench \cite{toolbench}, the prompt information describing the tool was generated dynamically from the reactor configuration. 
The model was then required to provide one or more tool calls to the reactor API required to carry out the growth experiments that fulfilled the query. These responses were encoded using JSON.

Each benchmark challenge required the model to choose the channels corresponding to the precursor and the co-reactant and the number of cycles. Some more complex challenges
required growing doped materials, where $n\times\mathrm{AB} + n\times\mathrm{CD}$ supercycles are used to form a doped or multicomponent material. A third type of challenge involved a functionalization step requiring only one precursor\cite{Yanguas-Gil2026aa}.

As in the ToolBench protocol, responses are scored based on how closely they match the target solution. Errors could be due to the model failing to provide the right JSON (corresponding to the failure to construct the right tool call response), selecting the wrong process (different channel numbers), or whether the right process was used but the wrong number of channels were provided. These two cases correspond to the case of using the wrong arguments. Furthermore, the study broke down the challenges in two broad categories: the first were instruction challenges, in which the agent was required to translate a request, for instance "Grow ten cycles of TMA/water" into the right function call given the reactor configuration. The second were process-discovery challenges, in which the query provided the desired outcome, for instance "Grow ten cycles of alumina", and the model had to infer both the right precursor and the channels. 

\begin{figure}[ht]
\centering
\includegraphics[width=15cm]{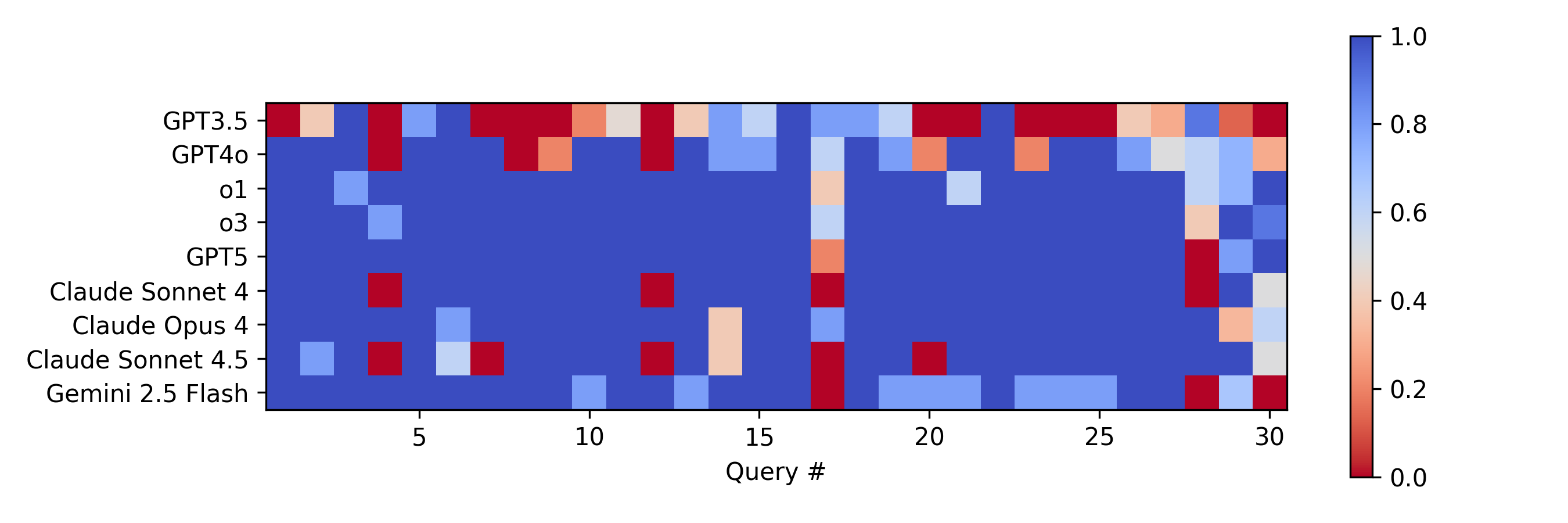}
\caption{Performance across 30 ALD process identification challenges of a simple agent built on different large language models. Each datapoint is the average of 5 independent runs. The scale is from 0 (complete failure) to 1 (perfect score). Reproduced from Ref. \cite{Yanguas-Gil2026aa} with permission. \label{fig:eval_aldtool}}
\end{figure}

In Figure \ref{fig:eval_aldtool} we show the performance for identical agents relying on different LLMs for the process discovery challenges and in Table \ref{tab:summary} we reproduce the key results. All scores are between 0 (bad) to 1 (perfect). The results show that, while the majority of the model excelled at instruction challenges, models comparatively struggled on process discovery challenges. This highlights the need to structure benchmarks in a way that highlights the model's performance in specific skills. The results in Table \ref{tab:summary} suggest that the models are good at translating direct instructions into function calls to an experimental tool, but that they find identifying the right processes more challenging. 

\begin{table}
\caption{\label{tab:summary}Average scores in instruction and process identification tasks as a function of the underlying large language model. The scale is from 0 (complete failure) to 1 (perfect score). Data from \cite{Yanguas-Gil2026aa}}
\begin{tabular}{lccc}
 Model & Instruction & Process Identification  \\
 \hline
 GPT3.5 & 0.80(0.09) & 0.39(0.04) \\
GPT4o & 1.0(0)	 & 0.72(0.04)  \\
o1 & 1.0(0) & 0.94(0.04)  \\
o3 & 1.0(0) & 0.96(0.03)  \\
GPT5 & 1.0(0) & 0.93(0.01)  \\
Claude Sonnet 4 & 1.0(0) & 0.85(0.02)  \\
Claude Opus 4 &1.0(0) & 0.93(0.02) \\
Claude Sonnet 4.5 & 0.96(0.05) & 0.78(0.02) \\
Gemini 2.5 Flash & 0.88(0.07) & 0.84(0.06)  \\
\hline
Average & 0.96(0.07) & 0.82(0.17)  \\
\end{tabular}
\end{table}

In fact, a more in-depth
analysis carried out in Ref \cite{Yanguas-Gil2026aa} showed that, on aggregate, models tend to perform worse on challenges involving processes that are not broadly represented in the literature as
well as those involving counter intuitive chemistries, such as the use of a typical oxidant as a co-reactant to grow a pure metal.  These insights were obtained because the process identification benchmark comprises 30 challenges tailored to a given materials synthesis technique. The results are also specific to the type of agent and the interface used
in that work. This highlights the importance of designing benchmarks that are specific to each experimental tool and interface and that can cover a wide range of different challenges and conditions.

\subsubsection{Closed-loop ALD process optimization benchmarks}
\label{sec:eval_aldopt}

Benchmarks focused on process optimization require the agent to interact with the experimental tool, similar to the multi-step manipulation scenario discussed in Section \ref{sec:eval_tool}. 
In the case of ALD,  this evaluation was done using a virtual tool, a simulation that, in addition to modeling an ALD process, returns feedback representative
of an in-situ characterization technique such as quartz crystal microbalance or spectroscopic ellipsometry, incorporating realistic constraints such as the addition of noise and a reduced number of significant digits.

While Paulson et al used this approach to evaluate various machine learning approaches\cite{Paulson2021}, Ref. \cite{Yanguasgil_reasoning} applied this methodology to evaluate the ability of agents built on top of reasoning large language models to optimize five different atomic layer deposition processes, chosen to represent a range of prototypical behaviors, including
differences in how rapidly saturation is reached, how sharp the saturation curve is, different values of the growth per cycle, and whether they had a non-self limited component. For most of the cases explored, the agent did not have any prior information about the ALD process. 

\begin{figure}[ht]
\centering
\includegraphics[width=7cm]{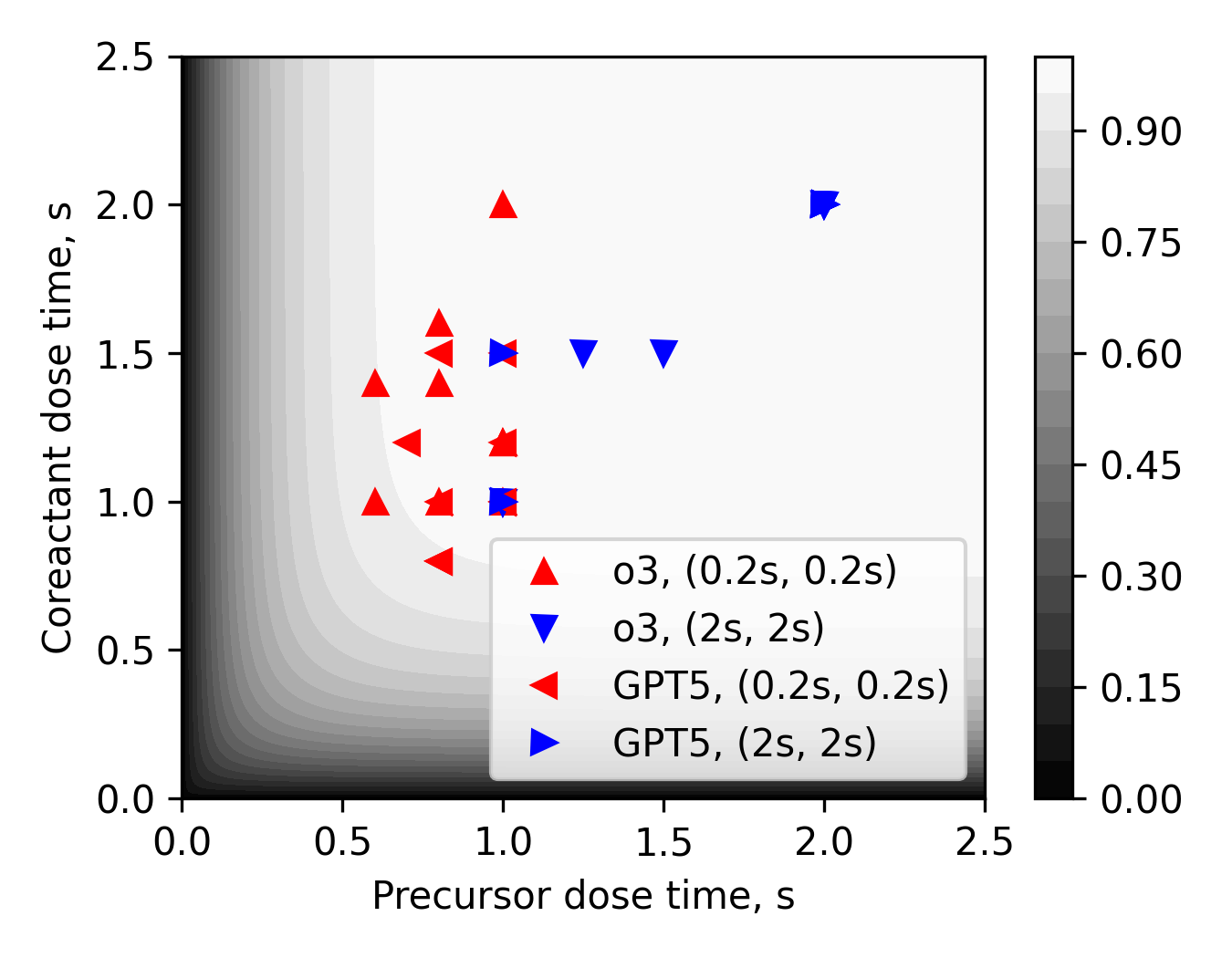}
\caption{Optimized dose time for one of the ALD process optimization baselines. In addition to the impact of the initial conditions, independent runs return different optimal conditions, highlighting the importance of considering the stochasticity of LLM's response. Fig. from Ref. \cite{Yanguasgil_reasoning} with permission.\label{fig:scatter}}
\end{figure}

As mentioned in Section \ref{sec:aldintro}, optimizing an ALD process involves identifying the shortest possible dose times that still yield a growth per cycle close to the saturated value. Consequently, the outcome of the optimization process is the final precursor and co-reactant dose times together with the corresponding growth per cycle. The total number of experiments requested and the number of iterations required by the agent to complete the optimization process were also tracked. These two numbers differ because the agent often requests more than one experiment per iteration.

\begin{figure}[ht]
\centering
\includegraphics[width=13cm]{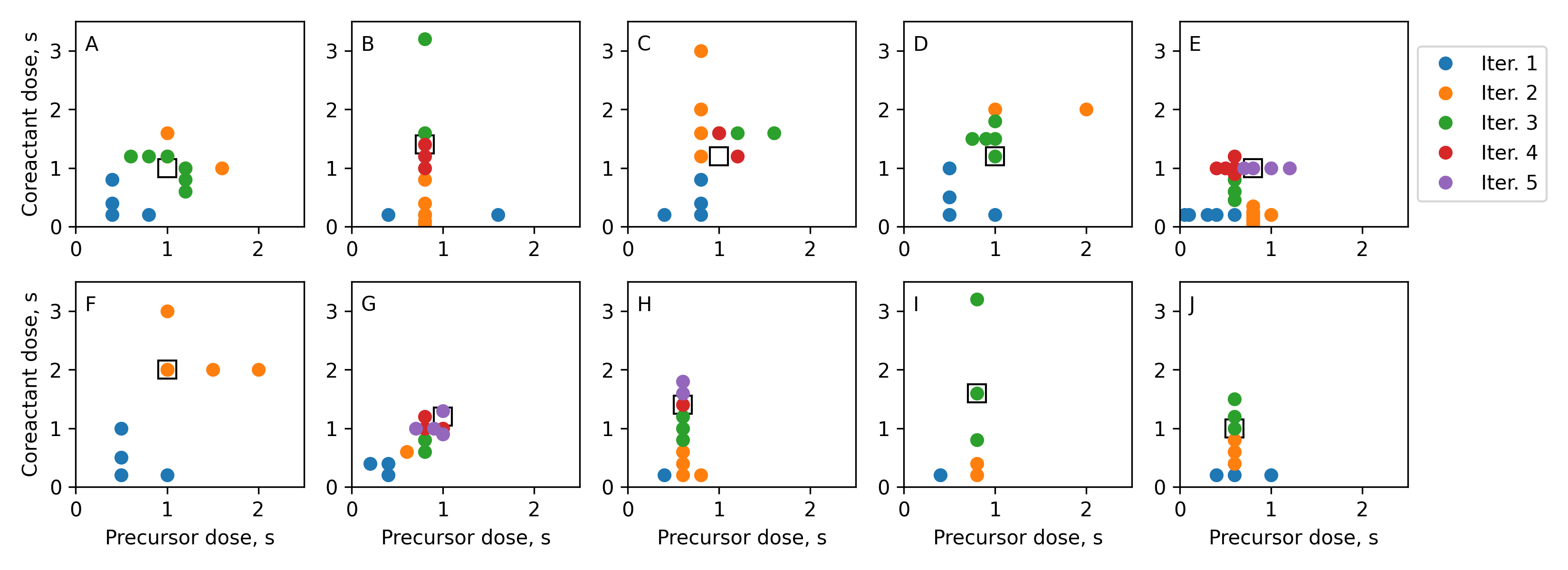}
\caption{Each plot in this figure represents an independent run of the agent trying to optimize the same ALD process. Points represent different conditions explored by the agent, with the color reflecting the iteration in which that specific growth was last requested. The plot shows a large variation in search strategies across ten different runs. Reproduced from Ref. \cite{Yanguasgil_reasoning} with permission. \label{fig:variability}}
\end{figure}

To address the inherent stochasticity of LLMs, ten independent runs were carried out per condition. While the agent almost always reported that the optimization had concluded satisfactorily, and the growth per cycle were close to the saturation growth per cycle, a significant variability in the dose times was observed even when starting from the same initial
condition. In Figure \ref{fig:scatter} we show an example of such variability in the optimization outcomes. Similarly, the search strategy changed from run to run. In Fig. \ref{fig:variability}, the explored conditions are colored by the iteration in which each growth was last requested.  These results highlight the importance of designing evaluation strategies that can evaluate the impact of stochasticity in the way an agent interacts with its environment.

\section{Discussion and practical recommendations}
\label{sec:discussion}

\subsection{A practical evaluation framework for materials synthesis}
\label{sec:eval_strategy}

As mentioned in Section \ref{sec:llm_eval},  there is a huge gap in our understanding of the performance of LLMs for materials synthesis. Only a few techniques have specific
benchmarks, and in most cases they have a limited coverage that does not provide a complete picture of the performance of LLM-based agents for that specific synthesis technique.
Even in the case of ALD treated in Section \ref{sec:ald}, the range of benchmarks is still incomplete. On the other hand, there are a number of well tested strategies to evaluate LLM-based models and agents that are directly applicable to materials synthesis.

This leads us to propose an evaluation strategy for materials synthesis, one that integrates different methodologies to evaluate LLM-based models and agents, including their integration with
experiments. This multistep approach is geared towards maximizing our understanding of the potential of these techniques in a way that is accessible to non-AI researchers:

\begin{enumerate}

\item {\bfseries Expert-reviewed open-response benchmark:} The first step is developing a set of questions to help gain a sense of the model's abilities and knowledge around a specific technique. The quality of the responses can be reviewed by domain experts, for instance using the methodology described in Section \ref{sec:aldbench}. From a methodological standpoint, the key consideration is to ensure that each question is answered independently. For instance, when using chatbot-type of interfaces, the model's context window grows with prior questions, which can affect the model's response. Using separate API calls for each question sidesteps this issue.

\item {\bfseries Multiple choice and open ended benchmark with LLM-as-a-judge evaluation:} Multiple choice questions and numerical open ended questions offer a scalable pathway to evaluate LLMs, particularly if LLMs are used as a judge to disambiguate open ended responses and extract any numerical value. This approach allows scaling up to thousands of questions, as shown
for the case of MatSciBench. The methodology should follow a workflow similar to that described in Figure \ref{fig:eval_workflow}. Given a corpus of research papers and other
background materials, approaches such as that of Li and Cole can be used to generate question sets using LLMs\cite{Li:2025aa}.

\item {\bfseries Low-level tool use benchmark:} These benchmarks should be directed at quantifying an agent's ability to generate one or more calls to the experimental tool, essentially the ability to translate instructions into concrete interactions with the experiment (e.g. "set the power of target 3 to 100 W for ten minutes"). These challenges can also be evaluated using
the same workflow described in Fig. \ref{fig:eval_workflow}.

\item{\bfseries Synthesis-challenge benchmark:} A variation of the previous case, it is possible to apply a similar methodology to materials synthesis challenges where, in addition to gauging the model's ability to execute specific instructions, it quantifies the model's ability to infer the instructions from the case. An example are the process identification challenges described in Section \ref{sec:eval_tool} for the case of ALD.

\item {\bfseries Closed-loop, multistep interaction benchmark:} This type of benchmark requires a virtual tool, a simulation or model of the experimental system capable of returning adequate feedback in a way that is consistent with specific characterization techniques, to allow the agent interact with the experimental system until a certain requirement is met. Since the goal is evaluating the performance of an agent in as general conditions as possible, virtual tools do not have to refer to any specific process or tool. Instead, they should be able to sample a wide enough range of behaviors to maximize the confidence on the agent's performance when facing a relevant research challenge in the lab.

\end{enumerate}

\subsection{Gaps, limitations, and opportunities}

A comparison of the state of the art with the evaluation strategy described in Section \ref{sec:eval_general} shows that there are huge gaps in our current understanding of the capabilities and limitations of models for materials synthesis. Even in the case of atomic layer deposition, arguably one of the best characterized techniques thus far, the picture is still far from complete. On the other hand, as emphasized in this review, over the past five years there has been a substantial development and refinement of evaluation methodologies for LLM models and agents\cite{Chang:2024aa}. There is therefore an opportunity to leverage those advances to develop benchmarks that are specifically tailored to different synthesis techniques.

One particularly important characteristic of LLMs is the stochastic nature of its output. In problems like ALD process optimization this leads to large run-to-run variabilities, as shown in Figs \ref{fig:scatter} and \ref{fig:variability}. Evaluation strategies targeting LLM-based systems need to address the impact of stochasticity if we want these techniques to become part of scientific research and materials synthesis workflows. A second challenge is understanding how robust the agent's response is to small perturbations in the way the question is phrased. We are currently working on simple tools to amplify existing materials science benchmarks to address this question. For experimentally integrated agents, evaluation should not only assess task success but also safe failure behavior, including the ability to interrupt, request clarifications, and avoid potentially harmful actions under uncertainty.

Finally, it is important to highlight that the specific LLM, generation hyperparameters, prompting strategy, tools and information accessible to the agent,
and the design of the harness itself will all affect its performance on any benchmark. Consequently, it is important to consistently keep evaluating the performance of any new agent on the relevant benchmarks to ensure that the performance does not regress as the design evolved. This ability to quickly iterate on an agent design is one of the main advantages of developing
a robust set of technique and tool specific benchmarks, providing a safe pathway towards fast design iterations without having to compromise on safety or having to wait for the outcome of the experiments they are controlling.

\section{Conclusions}

The integration of LLM-based agents with materials synthesis tools creates new opportunities for autonomous experimentation, but it also raises key questions in terms of
the performance, robustness, reproducibility, and safety of the AI models. In this review, we have provided an overview of evaluation strategies, and argued that a robust evaluation in the context of materials synthesis requires a careful approach spanning multiple facets, such as knowledge and reasoning, tool use, synthesis-specific challenges, and closed-loop interactions with experimental or virtual tools.

Atomic layer deposition provides a valuable case study because its high degree of automation enables evaluation at multiple levels, from process identification and tool use to full process optimization. The examples discussed here highlight both the promise of current LLM-based agents and their limitations, particularly with respect to their ability to infer new processes, their robustness, and their run-to-run variability. Developing more comprehensive, synthesis-specific benchmarks will be essential for safely integrating these systems into research workflows and accelerate the development of new models and agents that are tailored to materials synthesis needs.

\backmatter

\bmhead{Acknowledgements}

This research is based upon work supported by Laboratory Directed Research and Development (LDRD) funding from Argonne National Laboratory, provided by the Director, Office of Science, of the U.S. Department of Energy under Contract No. DE-AC02-06CH11357. The author also thanks the U.S. Department of Energy's Genesis Mission and the Office of Science, Office of Advanced Scientific Computing Research's ModCon.

\section*{Declarations}

\subsection*{Funding}

DE-AC02-06CH11357

\subsection*{Conflict of interest}

On behalf of all authors, the corresponding author states that there is no conflict of interest

\subsection*{Ethics approval and consent to participate}

Not applicable.

\subsection*{Consent for publication}

Not applicable.

\subsection*{Data availability}

All the data used to generate the figures is available from the author upon request.

\subsection*{Materials availability}

Not applicable.

\subsection*{Code availability}

Not applicable.

\subsection*{Author contribution}

The sole author responsible for the conception, literature review, and writing of this manuscript is AY

\bibliography{llm_paper}

\end{document}